\documentclass[utf8,final]{frontiersFPHY} 
\usepackage{url,hyperref,lineno,microtype,subcaption}
\usepackage[onehalfspacing]{setspace}

\newcommand{\um}{$\mu$m }
\newcommand{\uu}{$\mu$m}

\def\keyFont{\fontsize{8}{11}\helveticabold }
\def\firstAuthorLast{Hern\'an-Caballero {et~al.}} 
\def\Authors{Antonio Hern\'an-Caballero,$^{1,2,*}$, Evanthia Hatziminaoglou\,$^{2}$, Almudena Alonso-Herrero\,$^{3}$ and Silvia Mateos\,$^{4}$}
\def\Address{
$^{1}$Departamento de Astrof\'isica y CC. de la Atm\'osfera, Facultad de CC. F\'isicas, Universidad Complutense de Madrid, E-28040 Madrid, Spain\\
$^{2}$European Southern Observatory, Karl-Schwarzschild-Str. 2, 85748 Garching bei M\"unchen, Germany\\
$^{3}$Centro de Astrobiolog\'ia, (CAB, CSIC-INTA), ESAC Campus, E-28692 Villanueva de la Ca\~nada, Madrid, Spain\\
$^{4}$Instituto de F\'isica de Cantabria, CSIC-UC, Avenida de los Castros s/n, 39005, Santander, Spain
}

\def\corrAuthor{Antonio Hern\'an-Caballero}
\def\corrEmail{a.hernan@ucm.es}

\begin{document}
\onecolumn
\firstpage{1}

\title[Disk and dust in type 1 AGN]{Disentangling accretion disk and dust emissions in the infrared spectrum of type 1 AGN} 

\author[\firstAuthorLast ]{\Authors} 
\address{} 
\correspondance{} 

\extraAuth{}

\maketitle

\begin{abstract}

\section{}

We use a semi-empirical model to reproduce the 0.1--10\um spectral energy distribution (SED) of a sample of 85 luminous quasars.
In the model, the continuum emission from the accretion disk as well as the nebular lines are represented by a single empirical template (disk), where differences in the optical spectral index are reproduced by varying the amount of extinction. The near- and mid-infrared emission of the AGN-heated dust is modelled as the combination of two black-bodies (dust). 
The model fitting shows that the disk and dust components are remarkably uniform among individual quasars, with differences in the observed SED largely accounted for by varying levels of obscuration in the disk as well as differences in the relative luminosity of the disk and dust components. 
By combining the disk-subtracted SEDs of the 85 quasars, we generate a template for the 1--10\um emission of the AGN-heated dust. 
Additionally, we use a sample of local Seyfert 1 galaxies with full spectroscopic coverage in the 0.37\um to 39\um range to demonstrate a method for stitching together spectral segments obtained with different PSF and extraction apertures. We show that the disk and dust templates obtained from luminous quasars also reproduce the optical-to-mid-infrared spectra of local Seyfert 1s when the contribution from the host galaxy is properly subtracted.

\tiny
 \keyFont{ \section{Keywords:} galaxies:active, galaxies:Seyfert, quasars:general, infrared:galaxies, methods:data analysis, techniques: spectroscopic} %All article types: you may provide up to 8 keywords; at least 5 are mandatory.
\end{abstract}

\section{Obtaining accretion disk and dust templates from luminous quasars}

We use a sample of 85 luminous quasars ($\nu L_\nu$[3\uu]$>$10$^{45.5}$ erg s$^{-1}$) selected for their spectroscopic coverage (AKARI and/or Spitzer/IRS) in the rest-frame 2.5--10 \um range. In addition to the AKARI and Spitzer spectroscopy we obtain optical photometry from the Sloan Digital Sky Survey (SDSS) Data Release 12, near-infrared (NIR) photometry from the Two Micron All Sky Survey (2MASS), the VISTA Hemisphere Survey (VHS), and the UKIRT Infrared Deep Sky Survey (UKIDSS), and mid-infrared photometry in 4 bands from WISE \citep[see ][ for details]{Hernan-Caballero16b}.

In such luminous quasars the optical emission of the AGN easily outshines that of the host galaxy. The emission from dust is expected to be negligible at $\lambda$$\lesssim$0.85\um because the maximum temperature of dust grains is limited by sublimation to $\sim$1500 K \citep{Granato94}. As a consequence, only the accretion disk and the emission lines from the broad line region (BLR) and narrow line region (NLR) contribute significantly to the rest-frame 0.1--0.85\um spectrum. 

In \citet{Hernan-Caballero16b} we showed that a single empirical quasar template like that of \citet{Shen16} suffices to model the rest-frame UV-optical (0.1--0.85\uu) SED of luminous quasars if we allow for an adjustable extinction to reproduce the diversity in optical spectral indices among individual quasars.
To extend the template into the NIR and MIR ranges, we assumed that the disk emission follows a power-law with the theoretical slope $\alpha$=1/3 predicted for a locally heated optically thick disk \citep[e.g.][]{Shakura73,Hubeny01} and confirmed through polarized light observations \citep{Kishimoto08}. We also added to the new template the NIR nebular lines extracted from the quasar template of \citet{Glikman06}.

We fit the rest-frame 0.1--10\um SED of the 85 quasars with a two component disk+dust model. The disk component (plus nebular lines) is represented by the template described above, modified by an adjustable level of extinction with a wavelength dependency following the Small Magellanic Cloud Bar extinction law \citep{Gordon03}. This law is often used to de-redden quasars \citep[e.g.][]{Hopkins04,Glikman12} since it lacks the 2175 \AA{} absorption feature. The AGN-heated dust component is represented by the linear combination of two black-bodies (hot and warm) at adjustable temperatures within the intervals 850--2000 K and 150--900 K, respectively. 

\begin{figure*} 
\includegraphics[width=5.8cm]{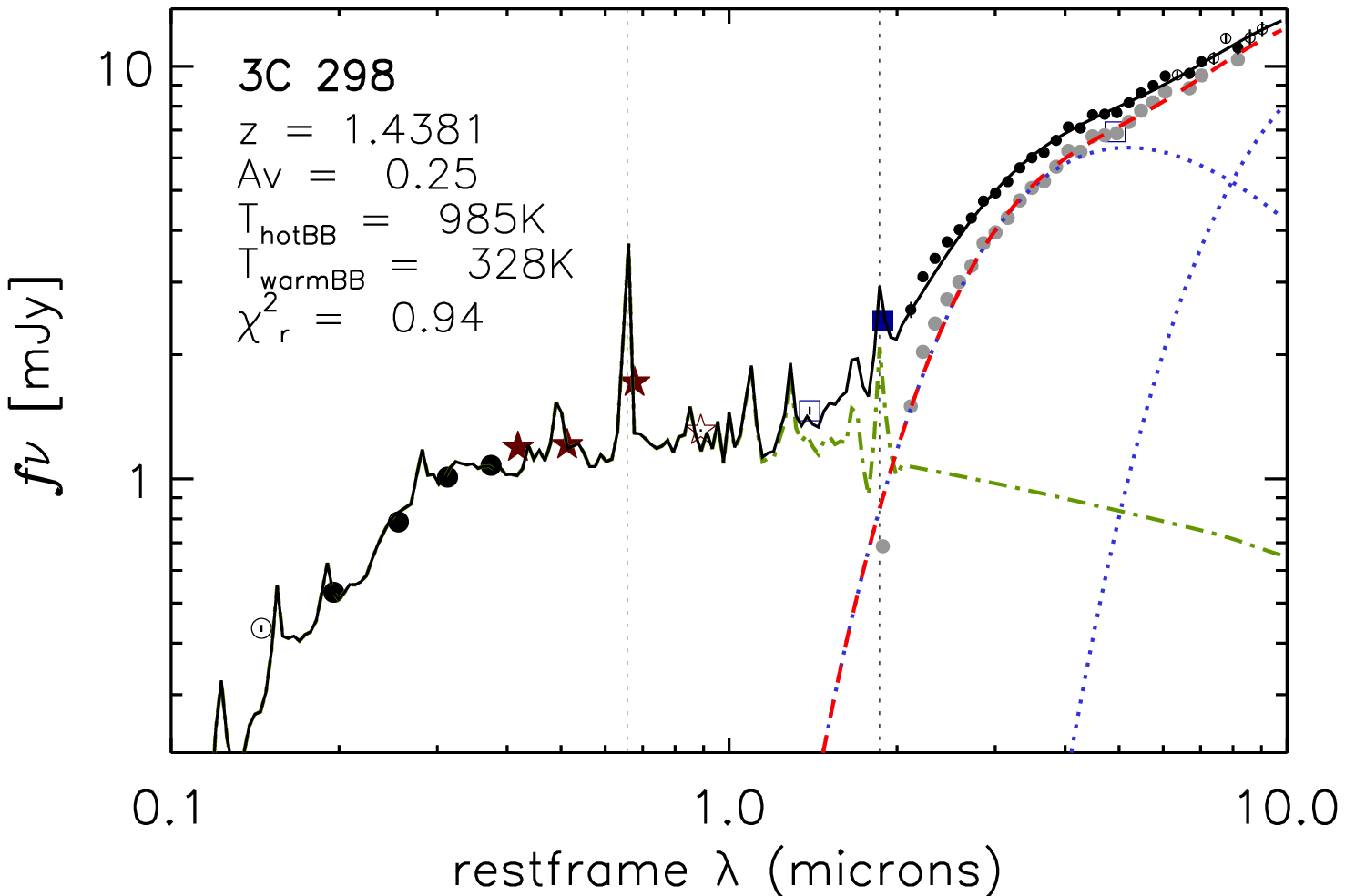}
\includegraphics[width=5.8cm]{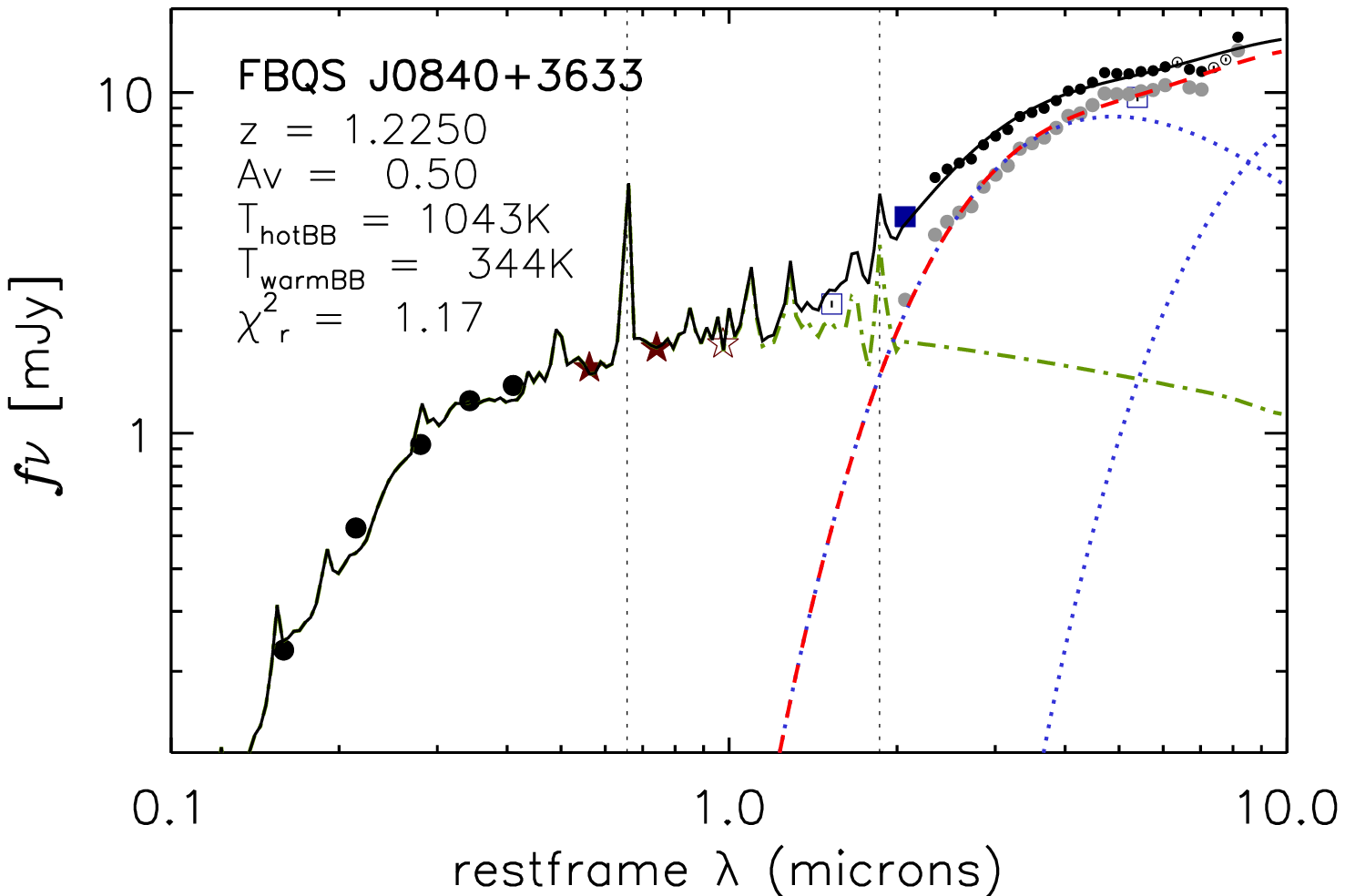}
\includegraphics[width=5.8cm]{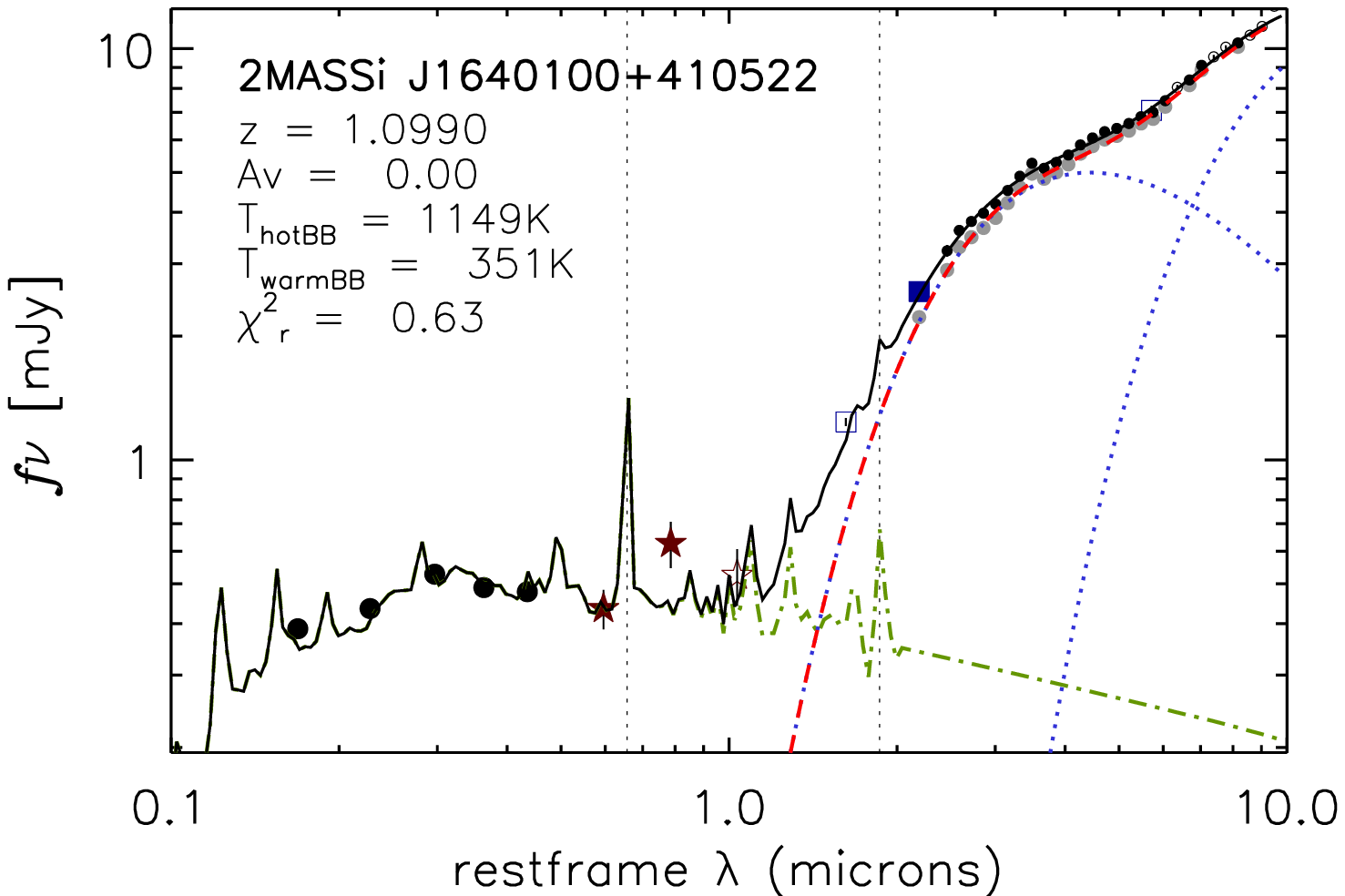}
\includegraphics[width=5.8cm]{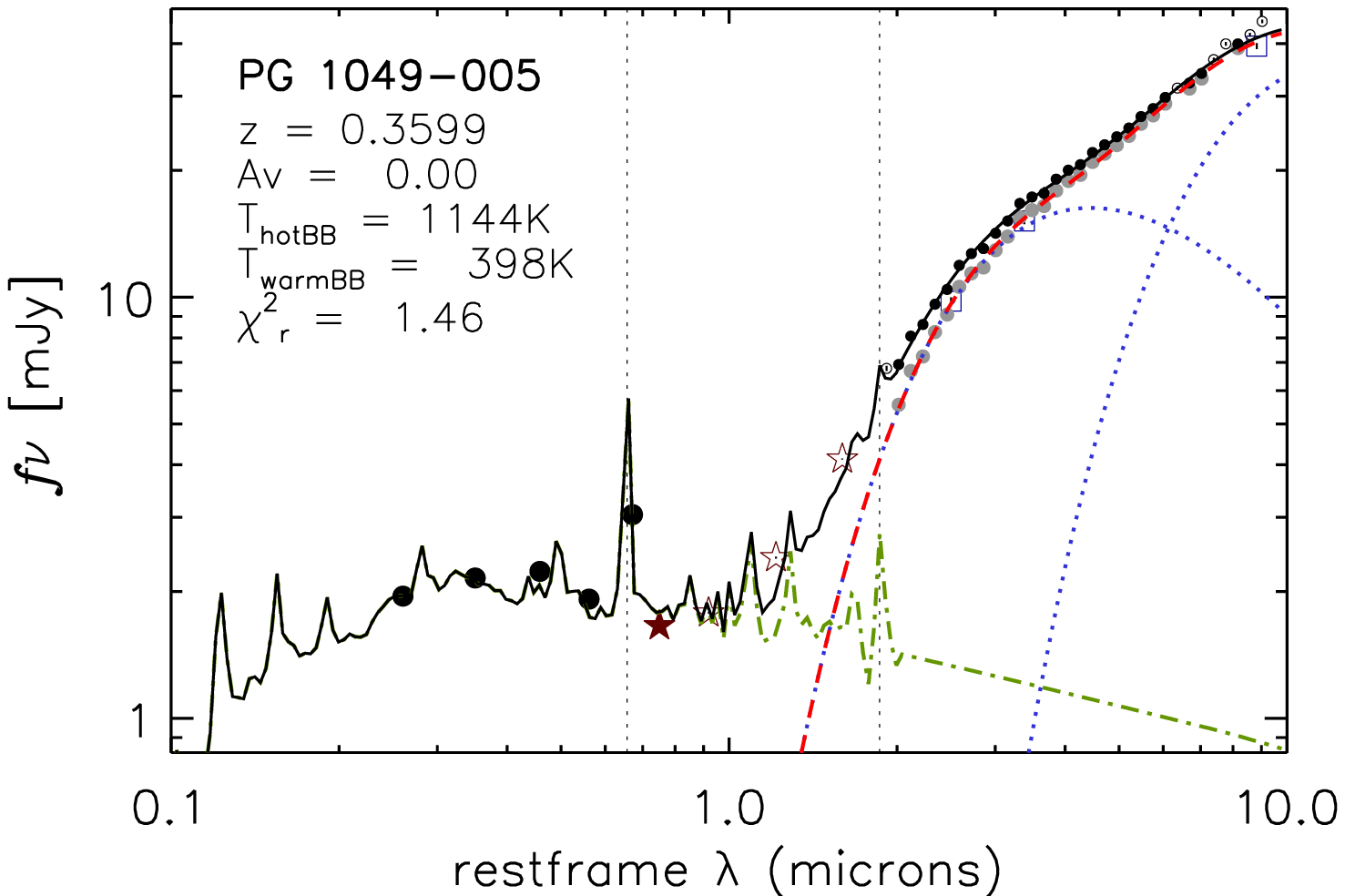}
\includegraphics[width=5.8cm]{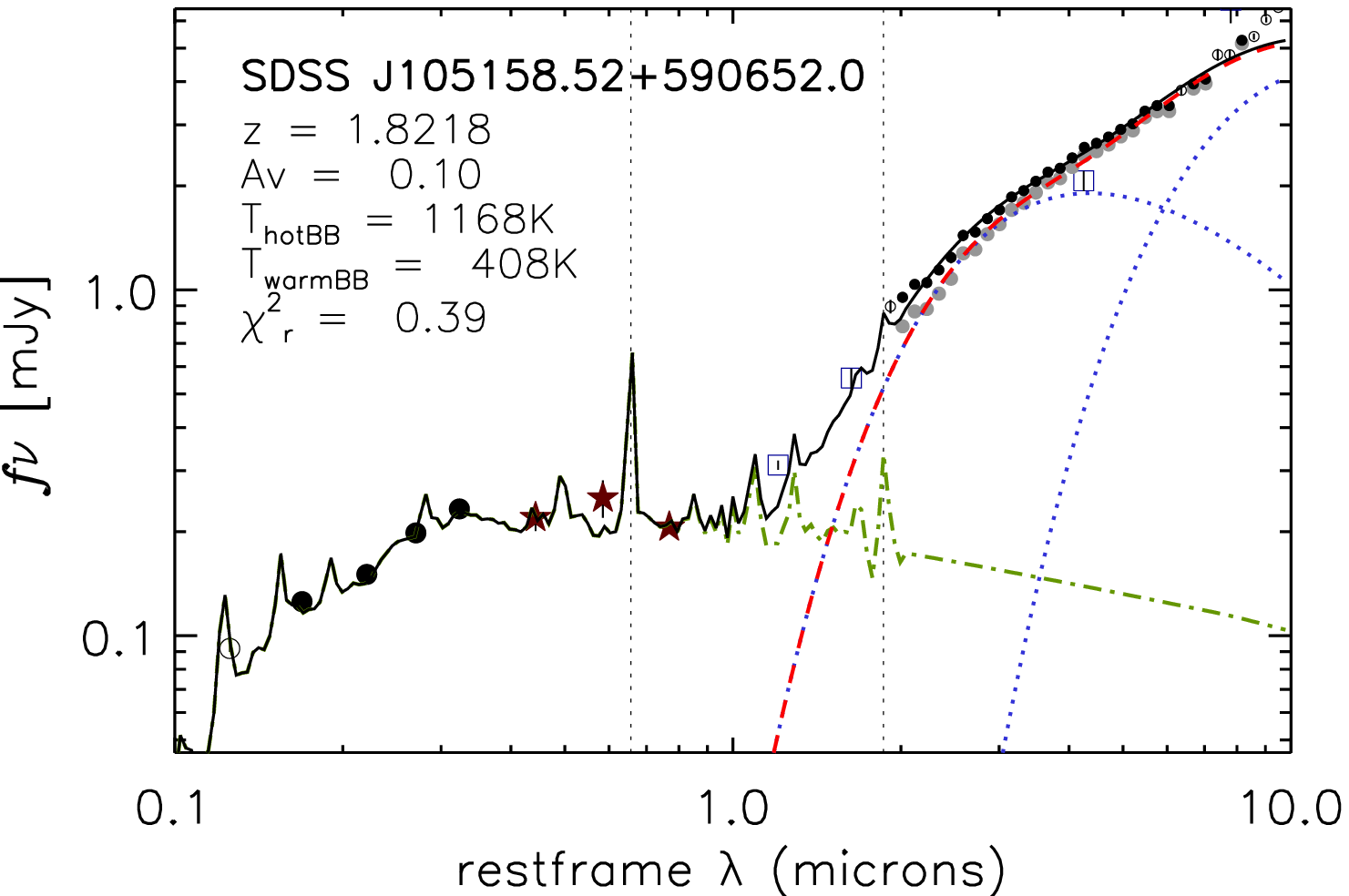}
\includegraphics[width=5.8cm]{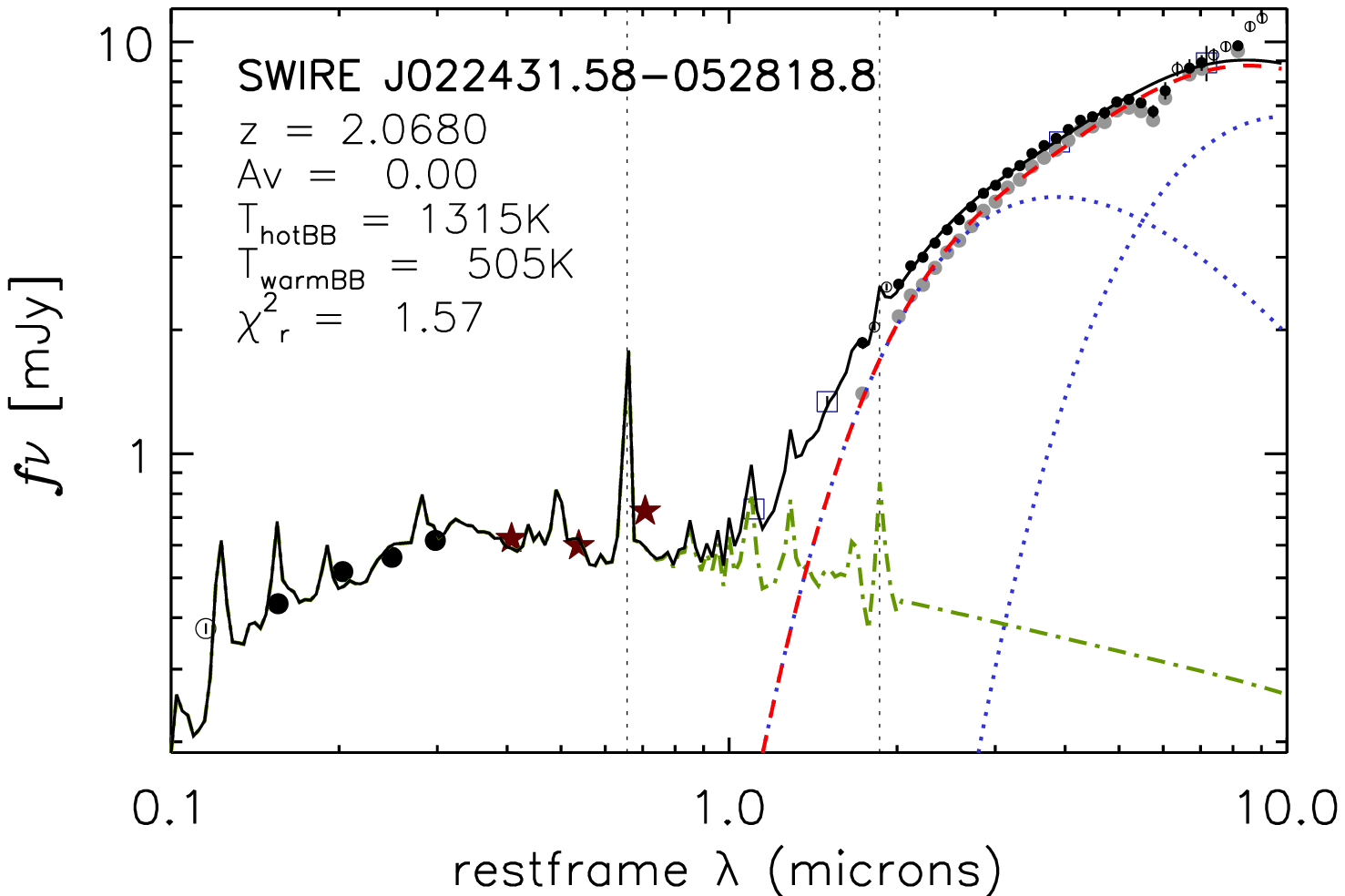}
\caption[]{Spectral energy distributions and their best fitting disk+dust models for a representative subsample of the luminous quasar sample showing a range of $A_V$ and dust-to-disk luminosity ratios. Circles, stars, and squares represent, respectively, broadband photometry in the observed-frame optical (from SDSS), NIR (2MASS/UKIDSS/VHS), and MIR (WISE). Open symbols indicate bands outside the wavelength range used to fit the disk or dust components. The disk model is shown as a green dot-dashed line. The (AKARI+)IRS spectrum resampled at $\Delta\lambda$/$\lambda$ = 0.05 is shown with small black dots. Grey dots below the (AKARI+)IRS spectrum represent the dust spectrum obtained after subtraction of the disk component. The model for the AGN-heated dust emission (red dashed line) is the linear combination of two black-bodies at adjustable temperatures (blue dotted lines). The combined disk+dust model is represented by the solid black line. The vertical dotted lines indicate the rest-frame wavelength of the recombination lines H$\alpha$ and Pa$\alpha$.\label{fig:fit-examples}}
\end{figure*}

Figure \ref{fig:fit-examples} shows examples of the best-fitting disk+dust decomposition model for a representative sub-sample of the 85 quasars. The residuals of the fit around $\sim$10\% are consistent with the photometric uncertainties. We find a systematic excess emission in the 1.0--1.5\um range relative to the model. The median, mean, and standard deviation of the excess at restframe 1.2\um is 32\%, 40\% and 33\%, respectively. This is comparable to the values found in \citet{Hernan-Caballero16b} in spite of the stronger NIR continuum and inclusion of NIR nebular lines in the new disk template, supporting our previous claim that the excess originates in the dust component.

\begin{figure}
\includegraphics[width=8.4cm]{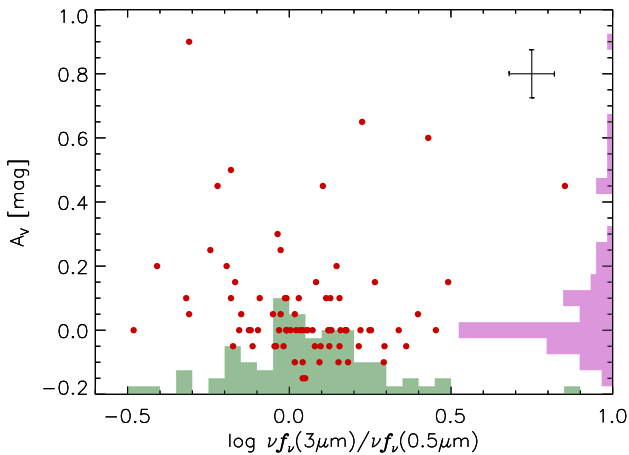}
\caption[]{Distribution of the extinction level applied to the disk component in the best fitting model and the dust-to-disk luminosity ratio for the whole sample of luminous quasars. Error bars in the top right corner indicate the typical uncertainties.\label{fig:Av-rNO}}
\end{figure}

The distributions of $A_V$ for the extinction of the accretion disk and the relative luminosity of the dust and disk components (represented by the luminosity ratio between $\lambda$=3.0\um and $\lambda$=0.5\uu) are shown in Figure \ref{fig:Av-rNO}. Negative values of $A_V$ are needed to model the disk of the bluest $\sim$25\% of quasars because the disk template is an average of observed spectra that have not been de-reddened. Therefore the extinction values obtained in the fit are relative to the (unknown) average extinction of the sample in \citet{Shen16}. Moderate values of $A_V$ (between -0.2 and 0.2 mag) fit all but the reddest 15\% of quasars. The $\nu L_\nu$(3\uu)/$\nu L_\nu$(0.5\uu) ratio peaks at $\sim$1, indicating that the peak luminosities of the disk and dust components are typically similar, albeit the full range of variation being a whole order of magnitude. 

\begin{figure}
\includegraphics[width=8.4cm]{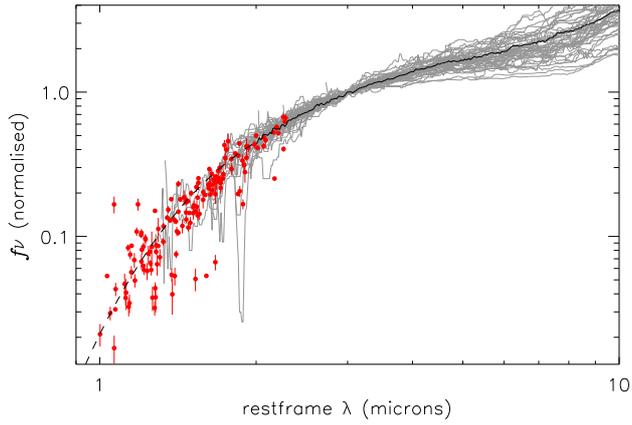}
\caption[]{Infrared SEDs of the individual quasars in the sample after subtraction of the disk component. The grey lines represent the (AKARI+)IRS spectrum, while the red dots represent the broadband data points at wavelengths with no spectroscopic coverage. All the individual SEDs are normalized at rest-frame 3\uu. The solid black line is the composite AKARI+IRS spectrum obtained for the same sample in \citet{Hernan-Caballero16b}. The dashed black line is an extrapolation of this composite using a T=1400 K black-body spectrum.\label{fig:nodisk}}
\end{figure}

Figure \ref{fig:nodisk} shows the infrared SEDs of the individual quasars, normalized at rest-frame 3\uu, after subtraction of the disk component. The SEDs show little dispersion ($<$0.3 dex) between $\sim$1.5\um and $\sim$6\uu, suggesting a largely invariable spectrum for the hot dust, that is consistent with black-body emission at a temperature close to the grain sublimation limit. The dispersion increases at longer wavelengths at least in part due to the onset of the broad silicate feature (which may be in emission or absorption). On the other hand, the larger dispersion at $\lambda$$\lesssim$1.5\um is a consequence of the uncertainty introduced by the subtraction of a disk component that is increasingly dominant at shorter wavelengths.

We obtain a template for the AGN-heated dust by averaging the AKARI+IRS spectra of the individual quasars. At $\lambda$$<$2\um only a few quasars have spectroscopic coverage, therefore we rely on the disk-subtracted broadband photometry to extend the template to $\lambda$$<$2\uu.
We find that a 1400 K black-body is consistent with most of the broadband data-points.

\section{Application to local Seyfert 1 galaxies}

To test whether the templates for disk and dust emission obtained for luminous quasars are also representative of the nuclear emission in less luminous type 1 AGN, we have performed spectral decomposition of a sample of 13 local ($z$$<$0.07) Seyferts and quasars with rest-frame 3\um luminosities in the range 10$^{42.8-44.4}$ erg s$^{-1}$.
The sample is a subset of the 23 broad-emission line AGN observed by \citet{Landt08}. 
We chose this sample because they obtained nearly simultaneous spectra in the optical (0.37--0.75\uu) and NIR (0.8--2.4\uu) with the FAST and SpeX spectrographs on the Tillinghast 1.5 m telescope and the NASA Infrared Telescope Facility, respectively. Out of their 23 sources, we select only the 13 that also have AKARI 2.5--5.0\um spectra from \citep{Kim15} and Spitzer/IRS 5.2--39\um spectra in the CASSIS \citep{Lebouteiller11,Lebouteiller15} or ATLAS-IRS \citep{Hernan-Caballero11} databases. Therefore we have continuous spectroscopic coverage from $\sim$0.35\um to $\sim$35\um in the restframe. However, since our templates are not defined beyond 10\uu, we perform the spectral decomposition only in the $\sim$0.35--10\um range.
 
Spectral decomposition on local type 1 AGN is more challenging compared to luminous quasars because the emission from the host galaxy is no longer negligible. The contribution from the stars to the total flux is important only at optical and NIR wavelengths, while the emission of the interstellar medium (ISM), in particular dust grains and aromatic hydrocarbons, is significant only in the mid-infrared.
Since the integrated spectrum of the stars in a galaxy peaks at 1.6\um and its NIR shape does not vary substantially among the different spectral types, we can use a single template to represent the stellar emission of the host galaxy in all the sources. We choose the S0 galaxy template from \citet{Polletta07}. In the optical, the actual spectra of the host galaxies may diverge from the template at shorter wavelengths depending on the age and extinction of the stellar population, but the difference only becomes important at $\lambda$$\lesssim$0.4\uu.

Another difficulty in decomposing local galaxies is that while the AGN emission is always spatially unresolved, the host galaxy is extended and its contribution to the observed spectrum depends on the spatial resolution of the observations and the size of the extraction aperture  (see \cite{Hernan-Caballero15} for a discussion).
The extraction aperture of the slit-less AKARI observations (60''$\times$7.5'') is wider than the IRS slit (3.6'' for SL1 and SL2 modules), and both are larger than those of the FAST (3'') and SpeX (0.8'') observations. To compensate for this, we compute the best-fitting disk+dust+stellar model for the local AGN as follows: for each of the four spectral segments (FAST, SpeX, AKARI, IRS) we obtain the values of the parameters that minimize the residuals while fitting by least squares the model:
\begin{equation}
F(\lambda) = a f_{disk}(\lambda) e^{-A_V \tau(\lambda)} + b f_{dust}(\lambda) + c f_{star}(\lambda)
\end{equation}
\noindent where $f_{disk}(\lambda)$, $f_{dust}(\lambda)$, and $f_{star}(\lambda)$ are the templates and $e^{-A_V \tau(\lambda)}$ is the extinction correction for the disk template. The coefficients $a$, $b$, and $A_V$ must be the same for all four spectral segments, but coefficient $c$ may take different values in each segment to compensate for the different amounts of host galaxy light in each aperture. 
We also apply a scaling factor between 0.7 and 1.3 to the fluxes in the FAST, AKARI and IRS segments to correct for any potential biases in the absolute flux calibration of up to $\sim$30\% relative to the SpeX spectrum. The values of these scaling factors are treated as free parameters in the fitting algorithm and they are computed independently for each source.

We use the coefficients in the best fitting model to obtain a stitched spectrum that merges the FAST, SpeX, AKARI, and IRS segments. For this we first multiply each segment (except the SpeX one, which is taken as reference) by its corresponding scaling factor, and then subtract the excess stellar component given by: 
\begin{equation}
\Delta F_{star}^{i}(\lambda) = (c^{i} - c^{SpeX}) f_{star}(\lambda)
\end{equation}
\noindent where the superscript $i$ indicates the spectral segment. The resulting stitched spectrum corresponds to our prediction of the spectrum that would be observed if all the segments had the same PSF, were extracted in the same aperture, and had perfectly consistent absolute calibrations. 

We evaluate the quality of the fits using the normalized root mean square error (RMSE), defined as:
\begin{equation}
RMSE = \sqrt{\frac{1}{N}\sum_i^N \frac{(f_i - F(\lambda_i))^2}{f_i^2}}
\end{equation}
\noindent where $f_i$ and $F(\lambda_i)$ are the flux densities at wavelength $\lambda_i$ in the stitched spectrum and the model, respectively. This statistic represents the typical relative residual between the stitched spectrum and the model, and is more informative than $\chi^2$ in situations where residuals are not dominated by noise but differences between the model and the intrinsic spectrum of the source \citep{Hernan-Caballero15,Hernan-Caballero12}.

\begin{figure*} 
\includegraphics[width=5.8cm,height=4.25cm]{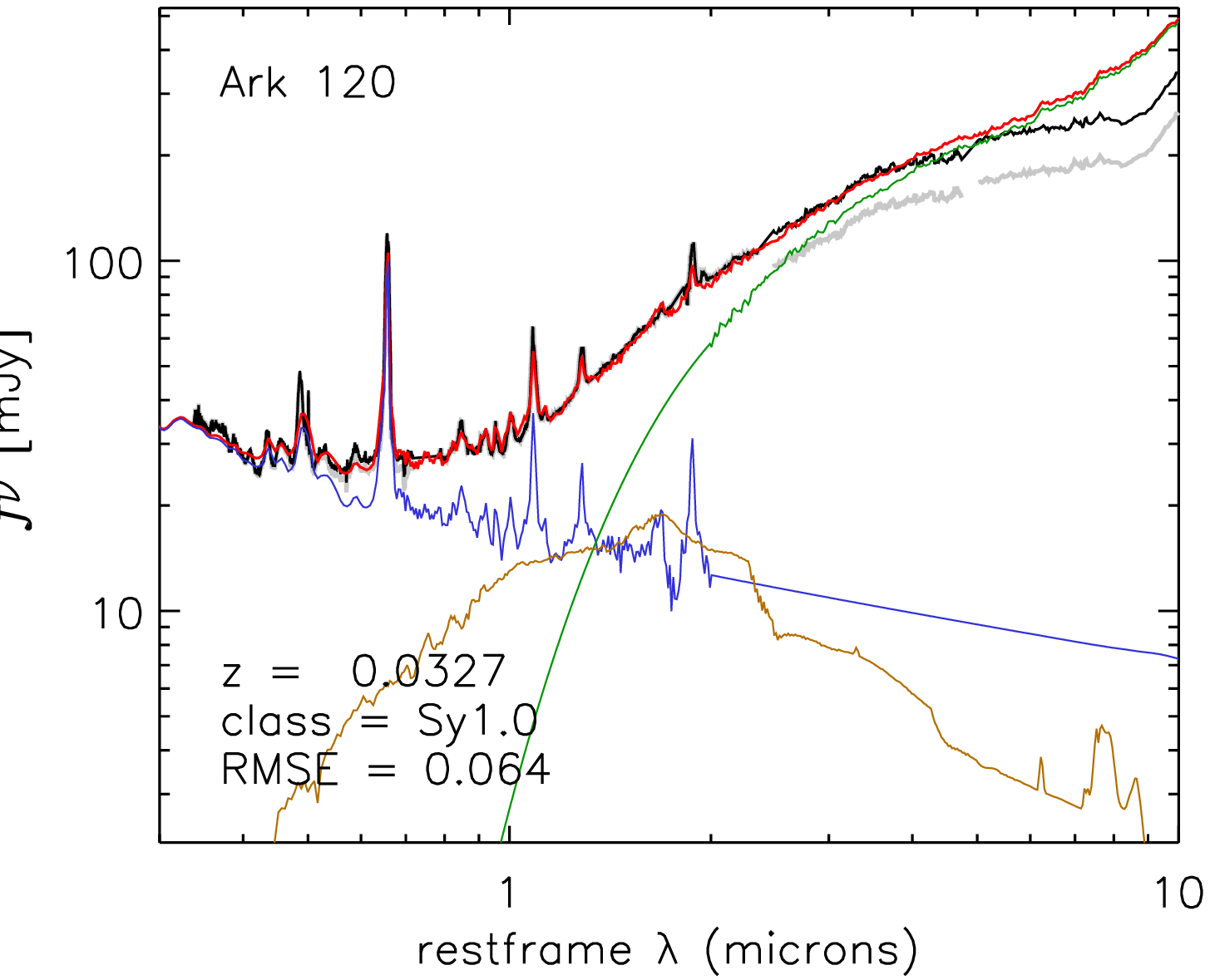}
\includegraphics[width=5.8cm,height=4.25cm]{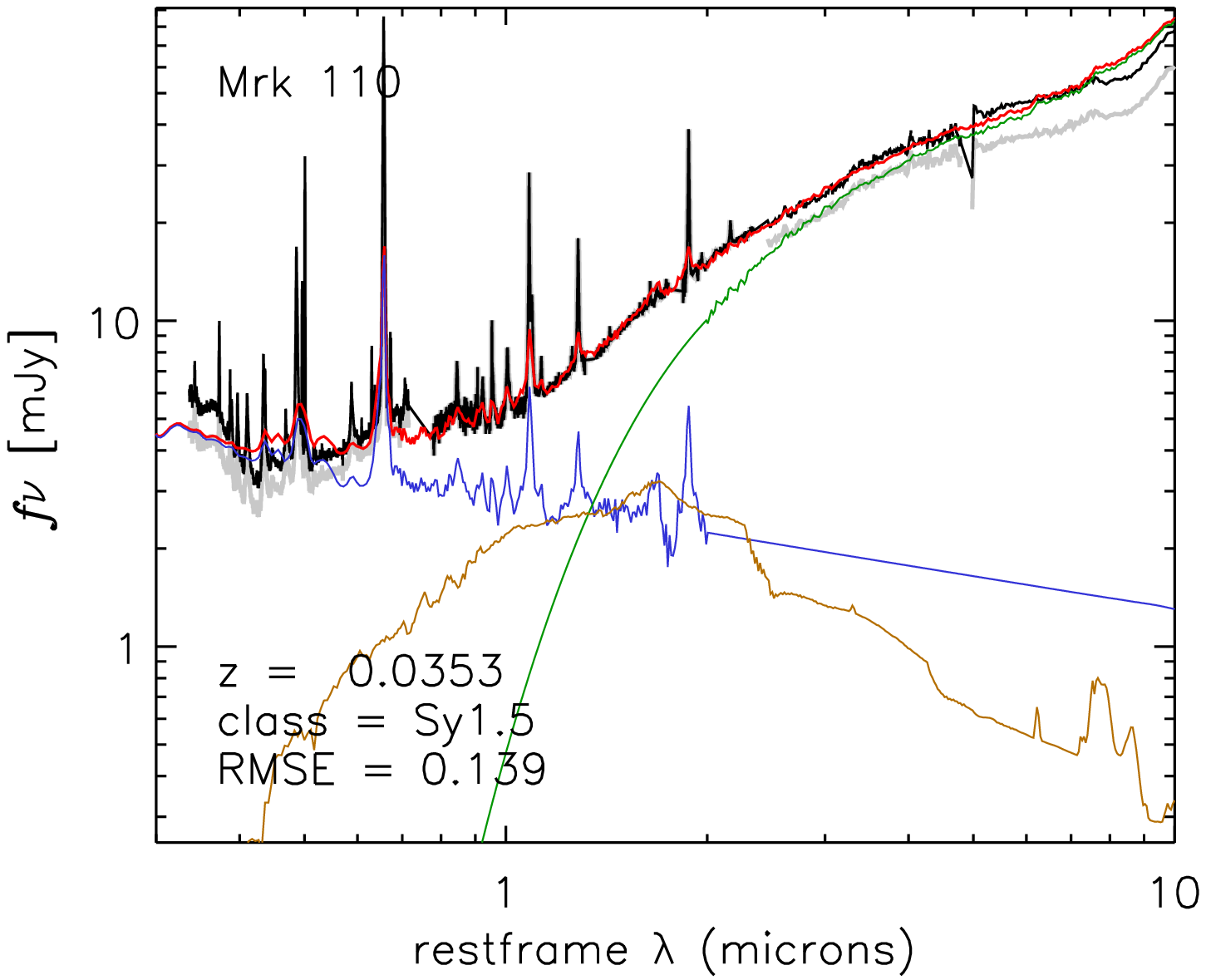} 
\includegraphics[width=5.8cm,height=4.25cm]{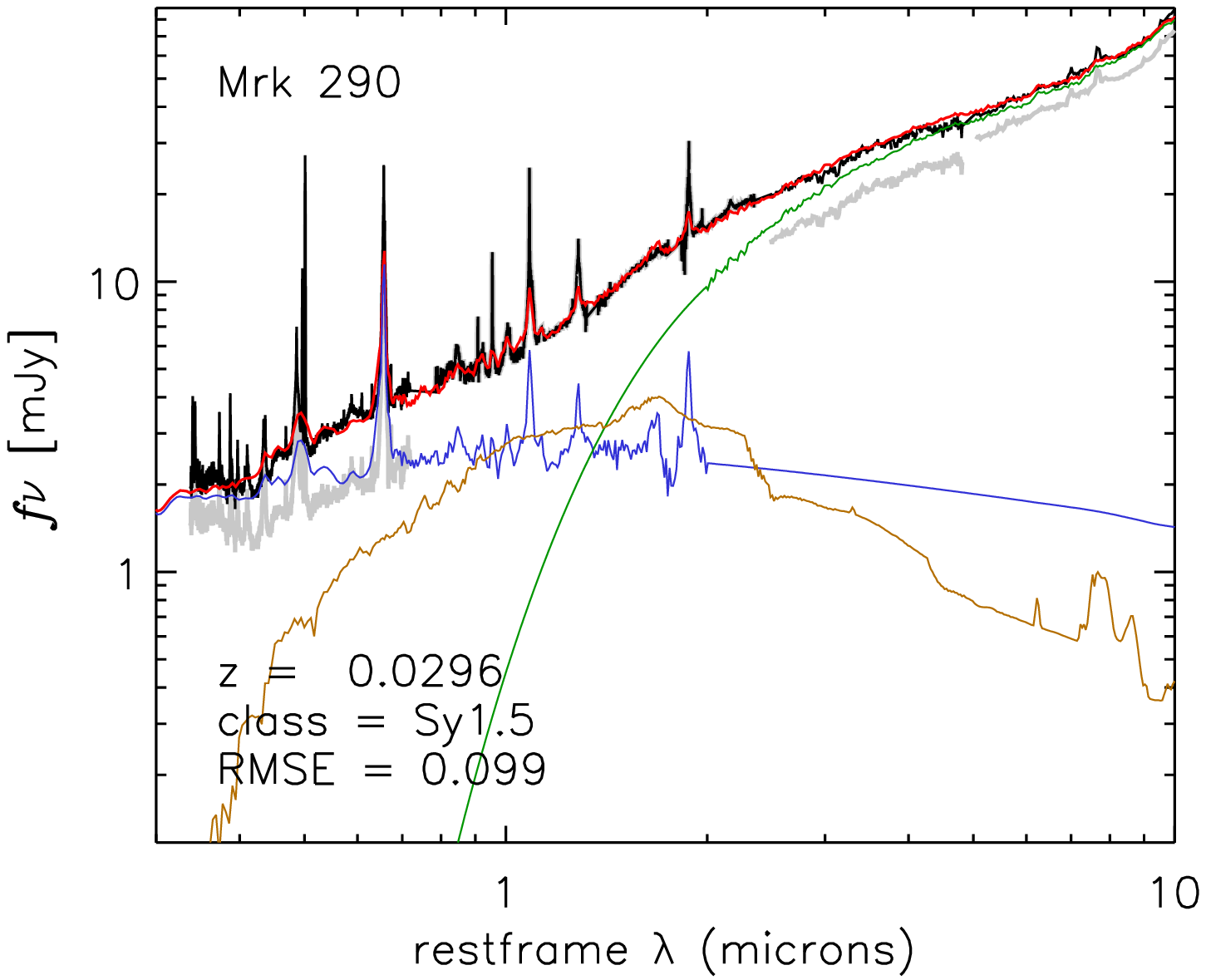}
\includegraphics[width=5.8cm,height=4.25cm]{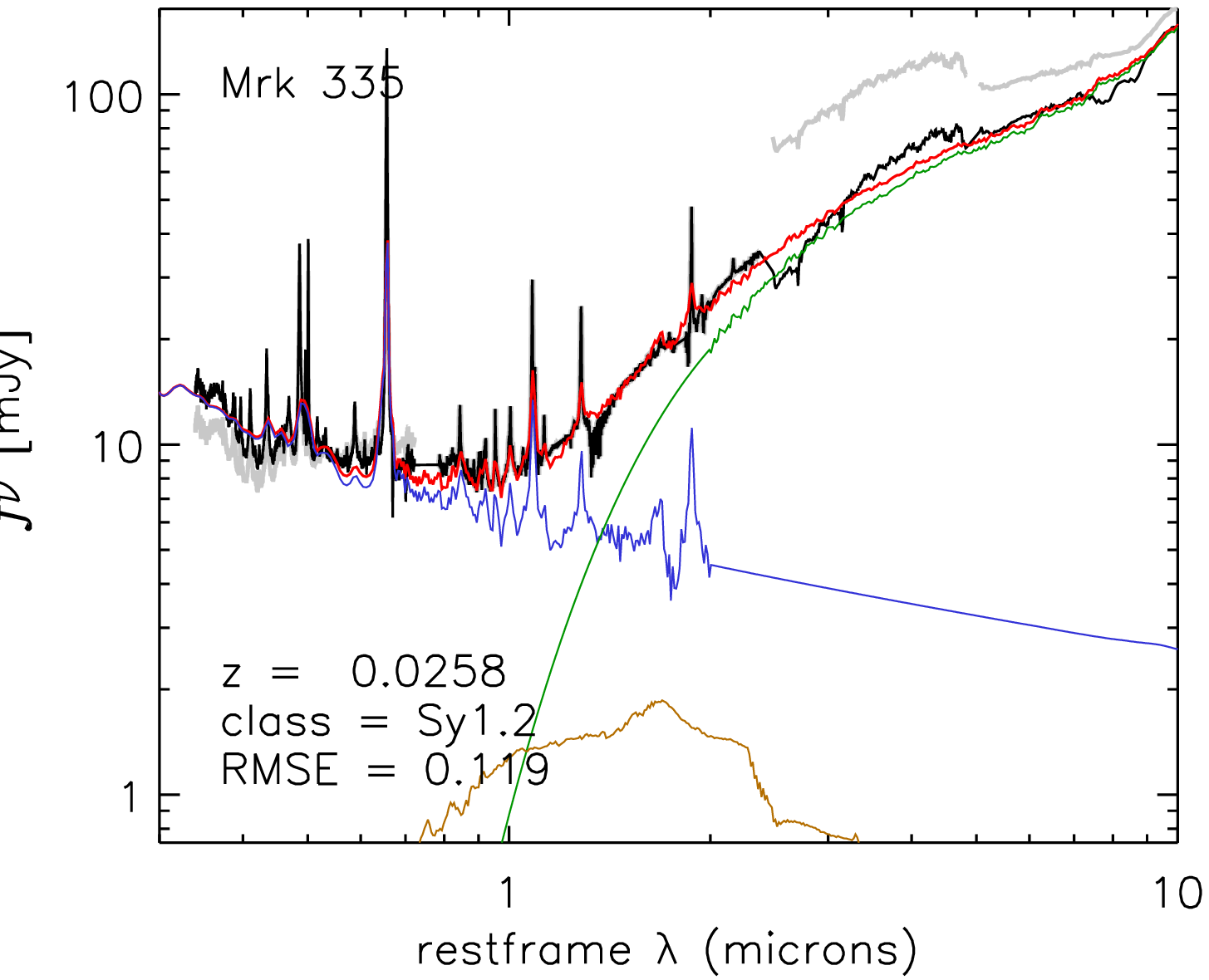}
\includegraphics[width=5.8cm,height=4.25cm]{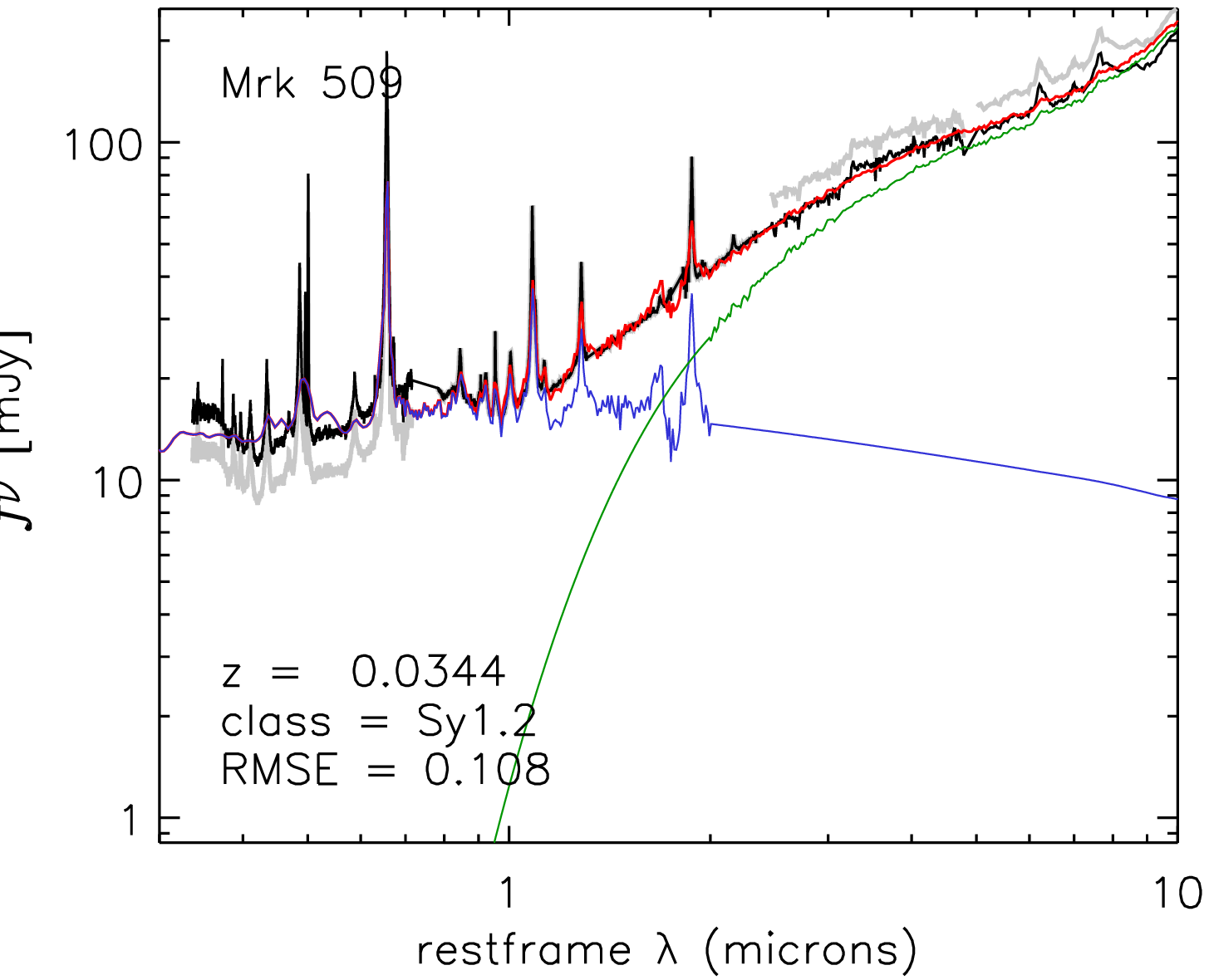}
\includegraphics[width=5.8cm,height=4.25cm]{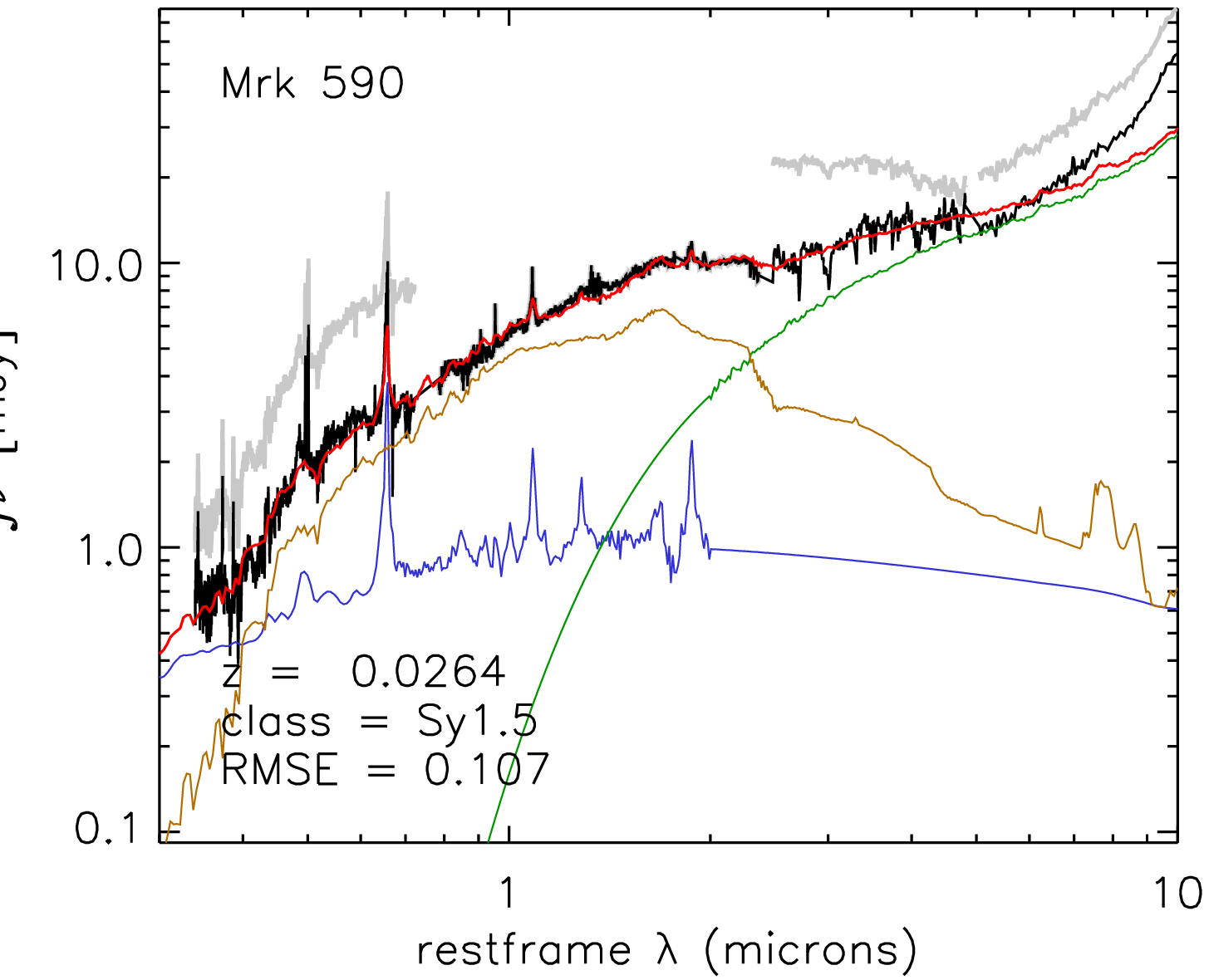}
\includegraphics[width=5.8cm,height=4.25cm]{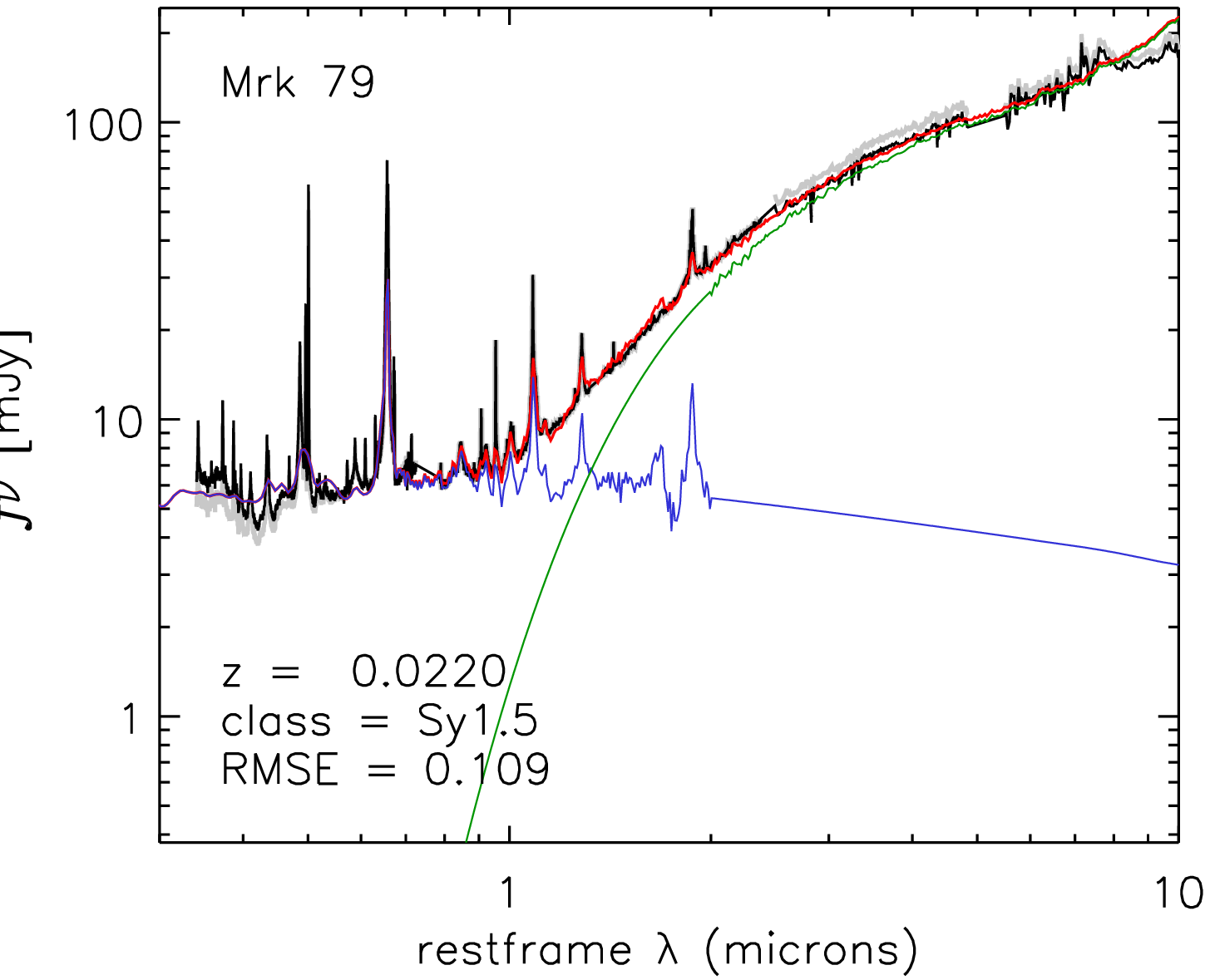}
\includegraphics[width=5.8cm,height=4.25cm]{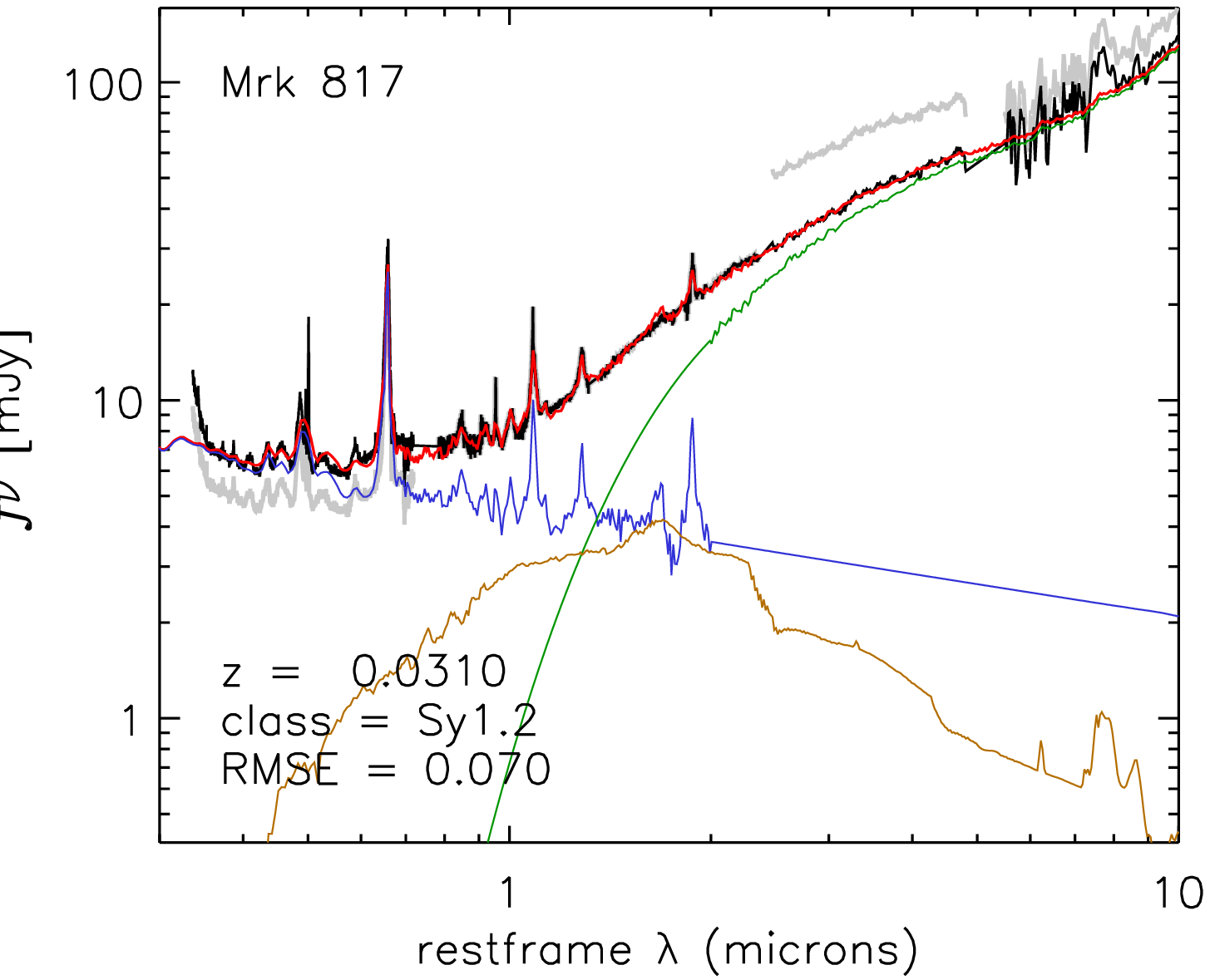} 
\includegraphics[width=5.8cm,height=4.25cm]{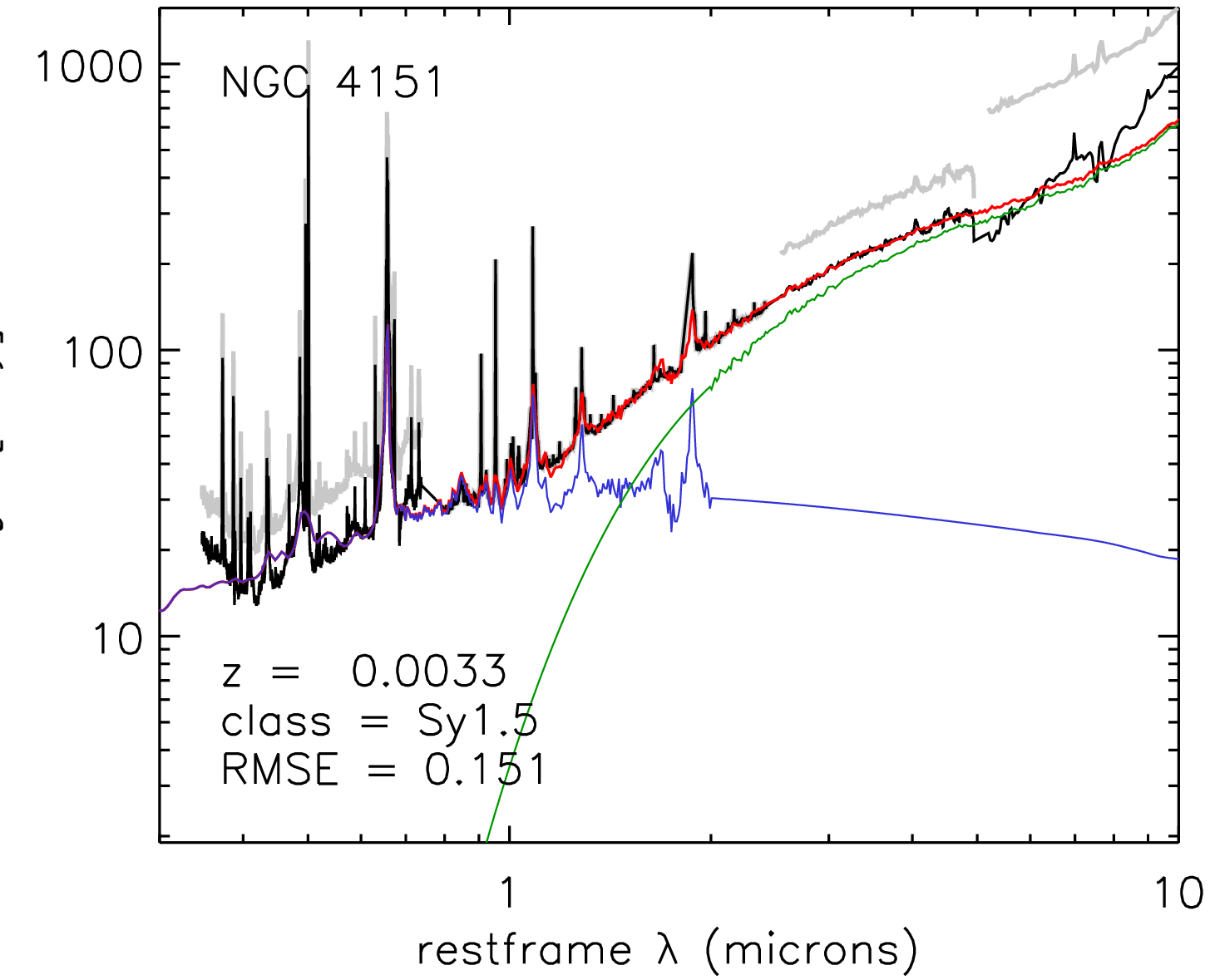} 
\includegraphics[width=5.8cm,height=4.25cm]{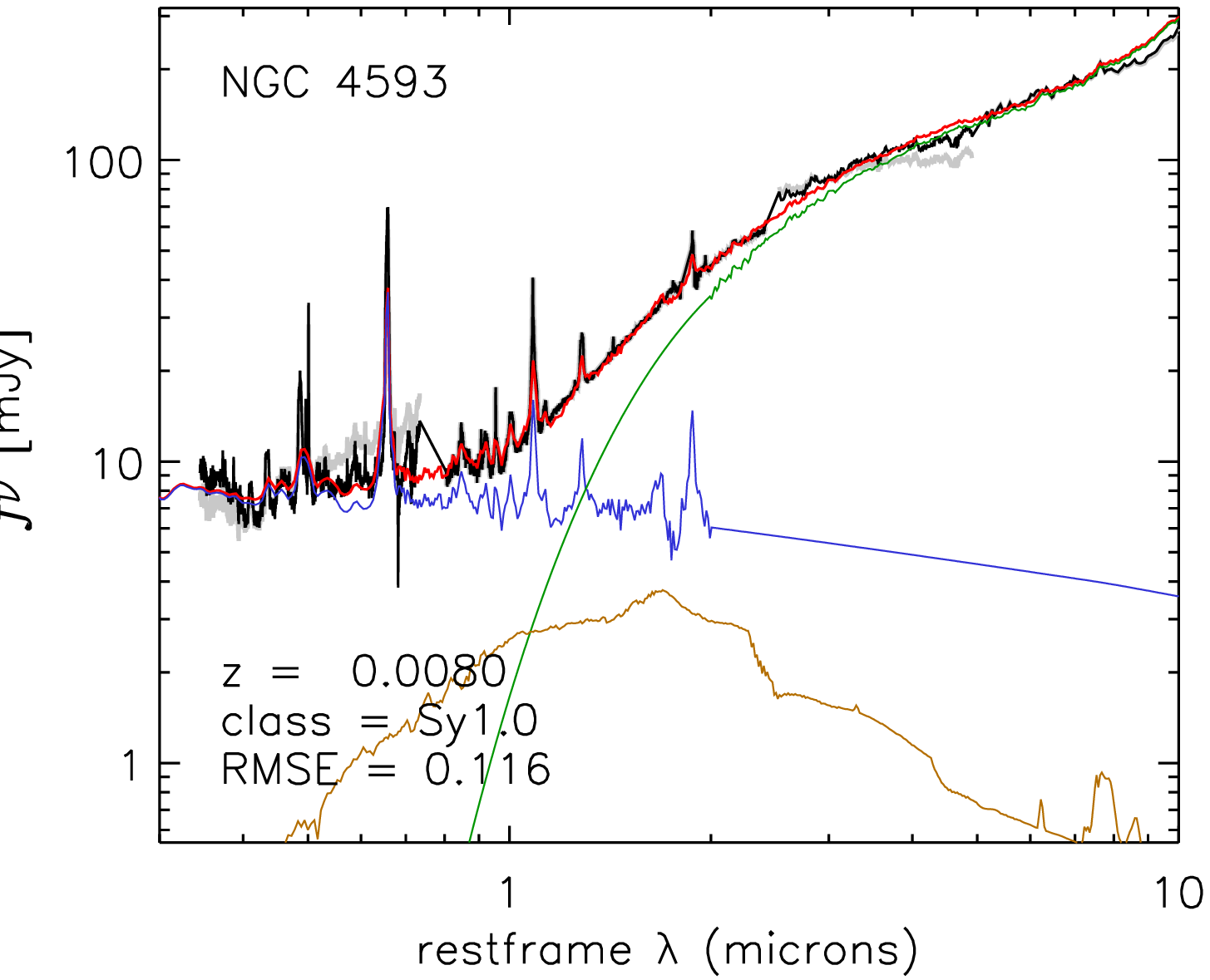}
\includegraphics[width=5.8cm,height=4.25cm]{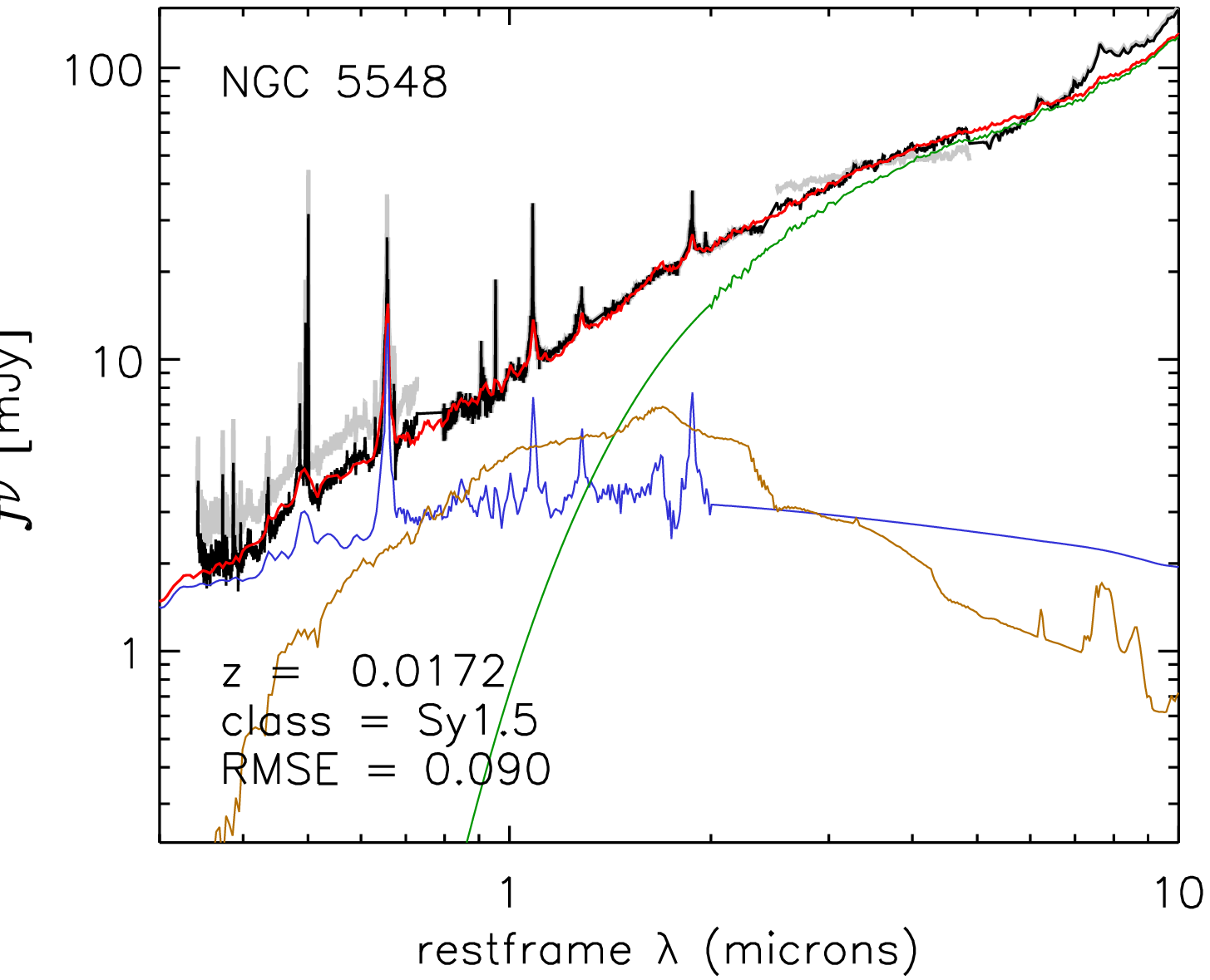}
\includegraphics[width=5.8cm,height=4.25cm]{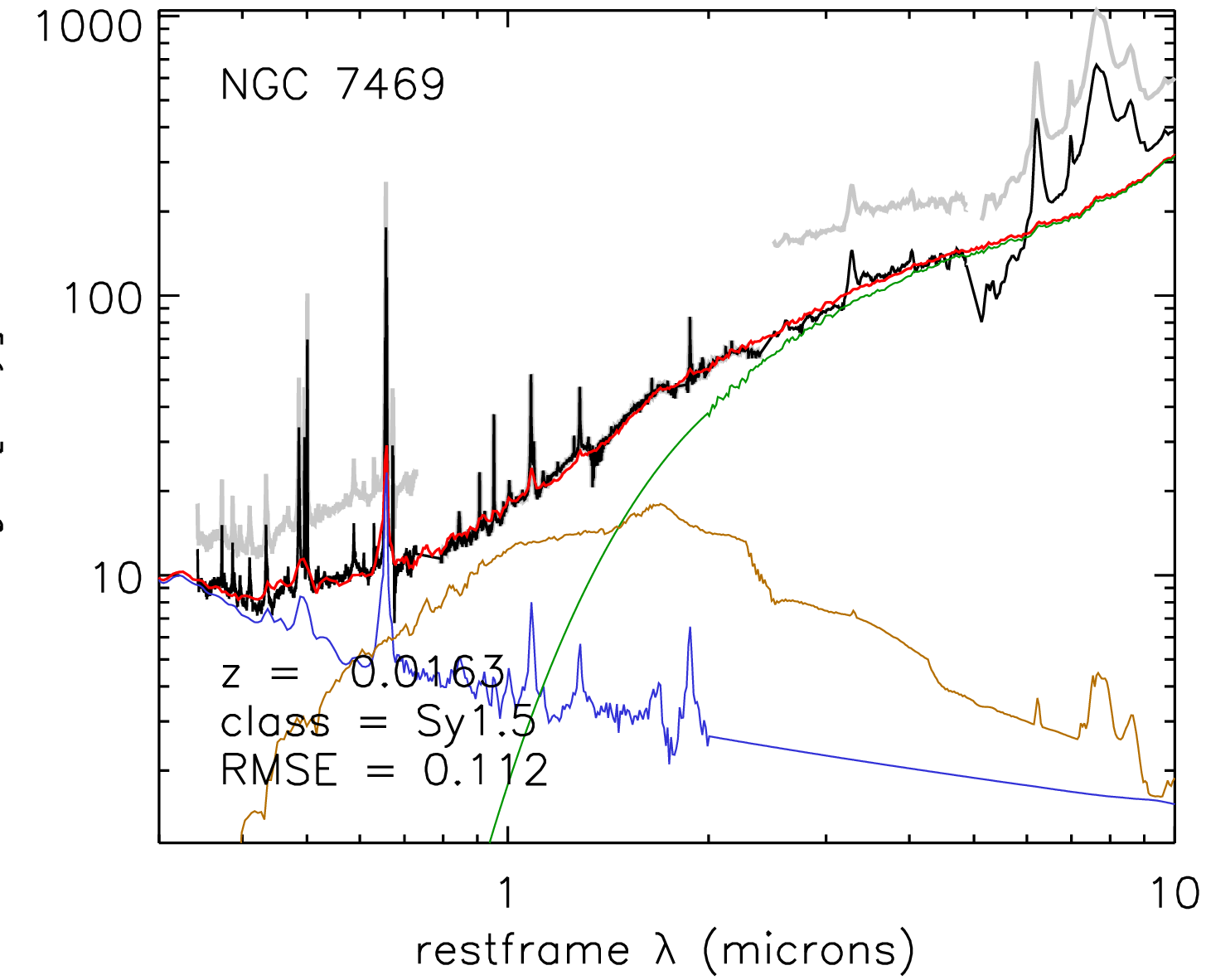} 
\includegraphics[width=5.8cm,height=4.25cm]{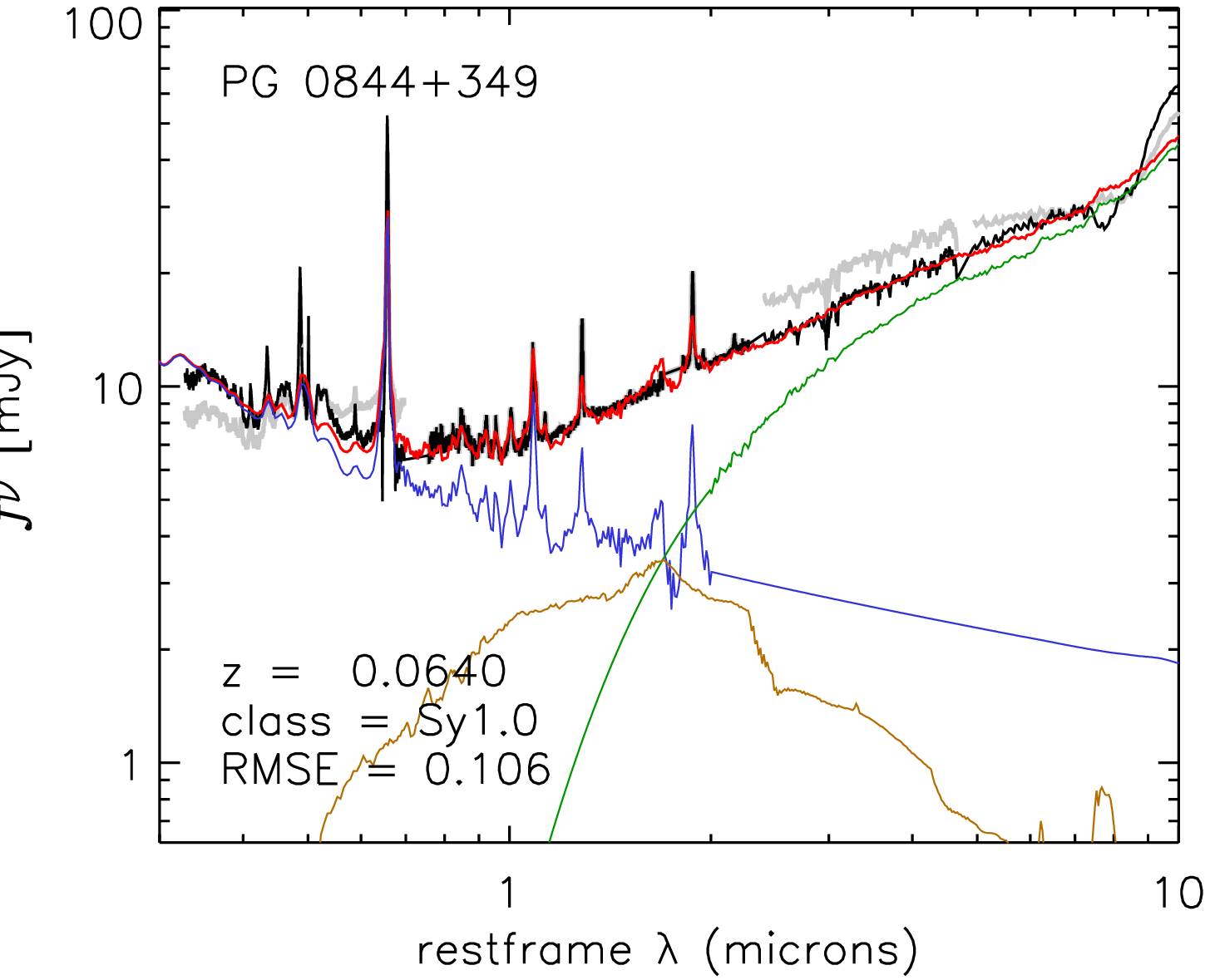}

\caption[]{Spectral decompositions of local type 1 AGN. The grey thick lines represent spectral segments from observations with FAST, SpeX, AKARI and IRS. The black line is the stitched spectrum. The red line is the best fitting disk+dust+stellar model for the stitched spectrum, while the blue, green, and yellow lines represent the individual disk, dust, and host galaxy components.\label{fig:fit-spectra}}
\end{figure*}

Figure \ref{fig:fit-spectra} shows the original and stitched spectra for the 13 sources in the sample, as well as the best fitting decomposition into disk, dust, and stellar components. 
The contribution from the stellar component to the total emission varies from negligible (e.g. Mrk 79, Mrk 509, and Mrk 335) to dominant (e.g. Mrk 590), while the interstellar emission is negligible in all but NGC 7469. There is no obvious correlation between the spectral class (Seyfert types 1.0, 1.2, and 1.5) and model parameters such as $A_V$ of the disk or the relative luminosities of the components, albeit the sample size is too small to draw any strong conclusions. 
The fits are remarkably good for most sources (typical RMSE $\sim$10\%), with significant discrepancies arising only at very short ($\lambda$$<$0.4\uu) and long ($\lambda$$>$7--8\uu) wavelengths. 
This suggests that the optical and NIR spectrum of the disk and dust emissions are relatively uniform among type 1 AGN regardless of luminosity. 
The discrepancies at short wavelength may be caused by uncertainty in the extinction affecting the disk component due to lack of restframe-UV observations, as well as increasing discrepancy between the assumed template for the stellar population and the actual spectrum of the host at short wavelengths.
Discrepancies at $\lambda$$>$7--8\um are caused by source-to-source variation in the strength of the mid-infrared aromatic features relative to the stellar continuum, and in some cases like Mrk 590, NGC 4151, and PG 0844+349, an unusually strong silicate emission feature at $\sim$10\uu. \citet{Hatziminaoglou15} showed that accurate modelling of the mid-infrared spectrum of Seyfert 1s requires a different approach, splitting the stellar and ISM emissions into separate spectral components and using multiple dust templates with different spectral indices and silicate strengths.

\section*{Funding}
This work was funded through the Spanish Ministry of Economy and Competitiveness (MINECO) grants AYA2015-70815-ERC, AYA2015-63650-P, and AYA2015-64346-C2-1-P.

\section*{Acknowledgments}
A.H.-C. acknowledges support from the ESO Scientific Visitor Programme.
This work is based on observations made with the \textit{Spitzer Space
Telescope}, which is operated by the Jet Propulsion Laboratory, Caltech
under NASA contract 1407.
The Cornell Atlas of \textit{Spitzer}/IRS Sources (CASSIS) is a product of the Infrared Science Center at Cornell University, supported by NASA and JPL.
This publication makes use of data products from the Two Micron All Sky Survey, which is a joint project of the University of Massachusetts and the Infrared Processing and Analysis Center/California Institute of Technology, funded by the National Aeronautics and Space Administration and the National Science Foundation.
This research has made use of the NASA/IPAC Infrared Science Archive, which is operated by the Jet Propulsion Laboratory, California Institute of Technology, under contract with the National Aeronautics and Space Administration.

\bibliographystyle{frontiersinHLTH&FPHY}

\begin{thebibliography}{999}
\bibitem[Hern\'an-Caballero et al.(2016b)]{Hernan-Caballero16b}Hern\'an-Caballero, A., Hatziminaoglou, E., Alonso-Herrero, A., Mateos, S., 2016, MNRAS,463, 2064
\bibitem[Granato \& Danese(1994)]{Granato94}Granato, G. L., Danese, L., 1994, MNRAS, 268, 235
\bibitem[Shen(2016)]{Shen16}Shen Y., 2016, ApJ, 817, 55
\bibitem[Shakura \& Sunyaev(1973)]{Shakura73}Shakura N. I., Sunyaev R. A., 1973, A\&A, 24, 337
\bibitem[Hubeny et al.(2001)]{Hubeny01}Hubeny I., Blaes O., Krolik J.~H., Agol E., 2001, ApJ, 559, 680
\bibitem[Kishimoto et al.(2008)]{Kishimoto08}Kishimoto, M., Antonucci, R., Blaes, O., et al. 2008, Nature, 454, 492
\bibitem[Glikman et al.(2006)]{Glikman06}Glikman E., Helfand D. J., White R. L., 2006, ApJ, 640, 579
\bibitem[Gordon et al.(2003)]{Gordon03}Gordon K. D., Clayton G. C., Misselt K. A., Landolt A. U. Wolff M. J., 2003, ApJ, 594, 279
\bibitem[Hopkins et al.(2004)]{Hopkins04}Hopkins P. F., et al. 2004, AJ, 128, 1112
\bibitem[Glikman et al.(2012)]{Glikman12}Glikman E., et al. 2012, ApJ, 757, 51
\bibitem[Landt et al.(2008)]{Landt08}Landt H., Bentz, M. C., Ward M. J., Elvis M., Peterson B. M., Korista K. T., Karovska M., 2008, ApJS, 174, 282
\bibitem[Kim et al.(2015)]{Kim15}Kim, D., et al., 2015, ApJSS, 216, 17
\bibitem[Lebouteiller et al.(2011)]{Lebouteiller11}Lebouteiller V., Barry D.J., Spoon H.W.W., Bernard-Salas J., Sloan G.C., Houck J.R., \& Weedman D., 2011, ApJS, 196, 8
\bibitem[Lebouteiller et al.(2015)]{Lebouteiller15}Lebouteiller V., Barry, D. J., Goes, C., Sloan G. C., Spoon, H. W. W., Weedman, D. W., Bernard-Salas, J., Houck, J. R., 2015, ApJSS, 218, 21
\bibitem[Hern\'an-Caballero \& Hatziminaoglou(2011)]{Hernan-Caballero11}Hern\'an-Caballero A., Hatziminaoglou E., 2011, MNRAS, 414, 500
\bibitem[Polletta et al.(2007)]{Polletta07}Polletta, M., et al. 2007, ApJ, 663, 81
\bibitem[Hern\'an-Caballero et al.(2015)]{Hernan-Caballero15}Hern\'an-Caballero A. et al., 2015, ApJ, 803, 109
\bibitem[Hern\'an-Caballero(2012)]{Hernan-Caballero12}Hern\'an-Caballero, A., 2012, MNRAS, 427, 816
\bibitem[Hatziminaoglou et al.(2015)]{Hatziminaoglou15}Hatziminaoglou E., Hern\'an-Caballero, A., Feltre, A., Pi\~nol-Ferrer, N., 2015, ApJ, 803, 110

\end{thebibliography}

\section*{Figure captions}

\end{document}